# Enhancing STEM Learning with ChatGPT and Bing Chat as Objects-to-Think-With: A Case Study

Marco Antonio Rodrigues Vasconcelos
PPGECIM - Doctoral Program in Science and Mathematics Education
ULBRA – Lutheran University of Brazil
E-mail: klavdianyves95@gmail.com

Renato P. dos Santos
CIAGE – Centre for Generative Artificial Intelligence in Cognition and Education
ULBRA – Lutheran University of Brazil
E-mail: RenatoPSantos@ulbra.edu.br

**Roles:**
RPdS contributed to the paper by providing ideas, formulating overarching research goals, and designing the methodology for the study. RPdS also oversaw the planning and execution of the research activity and provided mentorship to individual researchers. Additionally, RPdS applied formal techniques to analyse and synthesise study data. Finally, RPdS played a key role in preparing, creating, and presenting the published work, specifically writing the initial draft and providing substantive translation.
MARV played a key role in conducting the research and investigation process by performing experiments and collecting data and evidence. In addition, MARV made significant contributions to the preparation, creation, and presentation of the published work by revising the initial draft and providing valuable suggestions.

**Conflict of interest statement**
The authors have no conflicts of interest to declare. All co-authors have seen and agree with the manuscript's contents, and there is no financial interest to report. We certify that the submission is original work and is not under review at any other publication.

**Fundings**
No external funding was obtained for this project. The research was solely supported by the institutions of the authors.

# Enhancing STEM Learning with ChatGPT and Bing Chat as Objects-to-Think-With: A Case Study

**ABSTRACT**

This study investigates the potential of ChatGPT and Bing Chat, advanced conversational AIs, as "objects-to-think-with," resources that foster reflective and critical thinking, and concept comprehension in enhancing STEM education, using a constructionist theoretical framework. A single-case study methodology was used to analyse extensive interaction logs between students and both AI systems in simulated STEM learning experiences. The results highlight the ability of ChatGPT and Bing Chat to help learners develop reflective and critical thinking, creativity, problem-solving skills, and concept comprehension. However, integrating AIs with collaborative learning and other educational activities is crucial, as is addressing potential limitations like concerns about AI information accuracy and reliability of the AIs' information and diminished human interaction. The study concludes that ChatGPT and Bing Chat as objects-to-think-with offer promising avenues to revolutionise STEM education through a constructionist lens, fostering engagement in inclusive and accessible learning environments.

**CONTRIBUTION TO THE LITERATURE**

- This study fills a gap in current literature by investigating ChatGPT and Bing Chat's roles in enhancing STEM learning experiences.
- The application of ChatGPT and Bing Chat as objects-to-think-with, resources that foster reflective and critical thinking, problem-solving skills, and concept comprehension, is examined within the context of STEM education, offering a novel perspective to the field.
- Logs of students' extensive interactions with ChatGPT and Bing Chat are presented and analysed, providing valuable data on AI-assisted STEM learning experiences.
- ChatGPT's capacity to transform STEM education by fostering critical thinking, problem-solving, and engagement in inclusive and accessible learning environments is highlighted.

**INTRODUCTION**

The focus on developing critical thinking and collaborative skills has spurred interest in innovative approaches for enhancing education. Technology-based tools, especially in STEM education, are gaining attention. AI and NLP advancements have created opportunities for integrating technology into educational settings, particularly in STEM (Okonkwo & Ade-Ibijola, 2021).

Khosrawi-Rad et al. (2022) emphasize chatbots' educational advantages, such as constant availability and scalability. Kuhail et al. (2023) suggest chatbots can revolutionize education through engagement, personalization, educator support, and learner insight, particularly in STEM, where tailored support is beneficial.

In this context, generative AI-powered chatbots (GenAIbots), with OpenAI's ChatGPT emerging as a prime example and Microsoft's Bing Chat coming soon after, have recently garnered significant interest as educational tools in STEM learning (Baidoo-Anu & Owusu Ansah, 2023). Being capable of maintaining a convincing conversation with a user in natural language, those GenAIbots have emerged as promising tools to complement traditional pedagogical practices.

Probably, the most significant impact of ChatGPT is that no knowledge of any programming language is required; one 'talks' to it in one's native language, and it responds in one's – or any other language of one's choice! This is undoubtedly one of the main factors that made ChatGPT bring 100 million lay users to AI in just 2 months after launch (Milmo, 2023), with students making up a good portion of them.

OpenAI's ChatGPT[1], based on GPT NLP techniques, generates human-like responses using extensive internet text data (Radford et al., 2019). Microsoft's Bing Chat[2], built on GPT-4, is a revamp of traditional search engines, functioning as both search engine and chatbot (Mehdi, 2023). It accesses the internet directly, providing up-to-date, well-informed responses.

GPT-4 (OpenAI, 2023) showcases impressive language generation and user interaction capabilities, prompting investigation into using GenAIbots as objects-to-think-with in STEM education. Objects-to-think-with stimulate cognition, reflection, and understanding of complex concepts. This study examines GenAIbots' efficacy in enhancing STEM learning. As active agents, GenAIbots can shape learners'

---

[1] https://openai.com/
[2] https://www.bing.com/new

understanding, enabling exploration and reflective dialogues with ChatGPT and Bing chat, generating new insights. As will be seen later in this paper, those GenAIbots can work as instances of the concept of "object-to-think-with," resources that foster reflective and critical thinking, and concept comprehension, as introduced by Papert (1980, p. 11)

This paper showcases a case study that primarily investigates whether GenAIbots can effectively serve as objects-to-think-with, substantially enhancing students' engagement, knowledge acquisition, reflective thinking, and problem-solving abilities in STEM subjects.

The paper is organized as follows: First, a brief literature review discusses the relevance of GenAIbots in education, particularly in STEM learning. Next, the constructionist theoretical framework supporting this research is presented. This is followed by an outline of the case study research methodology, detailing the qualitative research design, participant selection, data collection instruments, and data analysis techniques. The subsequent sections present the study's findings, discussion, and conclusion, emphasizing the implications of GenAIbots as objects-to-think-with in enhancing STEM learning and identifying potential avenues for future research.

**LITERATURE REVIEW**

As AI-powered chatbots are a relatively new topic, there is limited literature available for a comprehensive review on this specific subject. Consequently, this study's literature review focuses on conventional chatbots, intending to derive valuable insights and foundational understanding that can be applied to AI-powered chatbots and their potential implications and benefits in educational contexts.

Okonkwo & Ade-Ibijola (2021) reviewed 53 articles on educational chatbots, discussing benefits, challenges, and future research. Benefits included information storage, personalized learning, and instant support, while limitations were ethical evaluation and user attitude. Kuhail et al. (2023) analyzed 36 papers, finding chatbots primarily web-based and utilized in various disciplines. Evaluations showed improved learning but also challenges in dataset training and usability. GPT-4 model-based chatbots, like ChatGPT and Bing Chat, have made significant progress in overcoming these challenges. Wollny et al. (2021) reviewed 74 publications, identifying research opportunities like broadening chatbot evaluations, examining learners' information needs, and investigating adaptability for educational purposes. This study addresses some of these points by evaluating ChatGPT and Bing Chat's cognitive impact and examining their adaptability in enhancing STEM education.

As illustrated in the next section, the theoretical frameworks proposed by Papert (1980) and Turkle (1984) offer valuable insights into GenAIbots' application in STEM learning.

**THEORETICAL FRAMEWORK**
**objects-to-think-with**

The foundation of our discussion is constructionism, a learning theory that emphasizes hands-on, experiential learning. This approach is especially relevant to STEM fields, where students are required to develop practical skills and apply theoretical concepts to real-world challenges.

A central concept in our discussion is "objects-to-think-with," as introduced by Papert, which refers to "any tool or resource that learners can use to explore and develop their understanding of a particular concept or domain" (1980, p. 11). The concept of objects-to-think-with is a way to describe how specific tools or objects can aid individuals in their learning and cognitive development, cultivate reflective and critical thinking, problem-solving skills, and concept comprehension, enabling the exploration of ideas and the development of novel thinking approaches (Papert, 1980). A wide range of items can serve as objects-to-think-with, including playing cards, photos, school maps, post-it notes, Lego bricks, cameras, smartphones, simulation software, and notably, GenAIbots.

Turkle (1984) discusses objects-to-think-with as cognitive tools that support problem-solving, critical thinking, and creativity. These cognitive artifacts link abstract and sensory knowledge between social and individual worlds (Turkle, 2007). According to Turkle (2007), "We think with the objects we love; we love the objects we think with." GenAIbots, as modern examples, provoke reflection on fundamental themes and make self-reflection enjoyable, attracting people to engage with them. In Papert's words, "thinking about thinking turns the child into an epistemologist, an experience not even shared by most adults." (Papert, 1980, p. 19). To paraphrase Turkle, GenAIbots are 'metacognitive machines' because they influence how we think about our own thinking (1984, p. 17).

Taylor et al. (2022) stress the need to recognize chatbots' agency in educational research, providing insights for enhancing practices. They argue that objects, including chatbots, actively shape our understanding of the world, aligning with posthumanist theories highlighting human and non-human entanglement (Haraway, 1990). Swan (2015) talks of a future world of productively coexisting multispecies intelligence, ranging from "classic" unenhanced humans to enhanced humans and many forms of generated AI.

Mollick and Mollick (2022) contend that GenAIbots can address classroom challenges such as promoting transfer, correcting overestimated comprehension, and fostering critical evaluation. However, students should use caution and critical thinking with GenAIbots, as they can 'hallucinate,' i.e., fabricate facts and references (OpenAI, 2023, p. 6), or give incorrect answers to mathematical computations. They are not substitutes for Google, calculators, or Alexa, and seeking justifications for incorrect responses may result in false explanations.

However, as Mishra et al. (2023) explain, the process of giving plain language instructions to a language model like GPT, called *prompting*, can be challenging for everyday users to create perfect prompts for consistent results. Mollick (2023) emphasizes that the best way to learn is to use AI as a partner and a teacher, experimenting with different prompts and engaging in a dialogue with the AI, asking it questions, giving it feedback, and challenging it to improve. Nevertheless, this process of successive reflections and interactions is also part of the use of GenAIbots as objects-to-think-with.

Chatbots' efficacy in generating educational natural language responses is well-established (Okonkwo & Ade-Ibijola, 2021; Sabzalieva & Valentini, 2023). ChatGPT excels at creating human-like responses (Franciscu, 2023), showing potential for STEM learning applications. GenAIbots facilitate personalized, adaptive, and collaborative learning (Dillenbourg, 1999, p. 1), encouraging diverse students to share ideas and learn together. They promote active learning, generating STEM examples and practice problems, enhancing understanding of intricated STEM concepts.

Consequently, incorporating GenAIbots as objects-to-think-with in STEM education can create a more inclusive, creative, and collaborative learning environment that promotes exploration and innovation. Utilizing GenAIbots enables learners to engage with complex concepts, gain new insights, reflect on their understanding, ask questions, receive instant feedback, and explore diverse perspectives.

**MATERIALS AND METHODS**

This section delineates the research methodology adopted for this case study. This investigation aims to discern the extent to which GenAIbots, a sophisticated language model, can augment STEM learning when utilised as an object-to-think-with. To ensure a rigorous examination, we have employed a qualitative research design, focusing on an in-depth exploration of participants' experiences and perceptions.

To comprehensively understand the students' interactions with GenAIbots as objects-to-think-with in STEM learning, this study employed an exploratory holistic single case study research design with two embedded units, namely ChatGPT and Bing Chat. This methodology enables the analysis of the interactions between these units, following the principles Yin (2011) outlined. The chosen approach allows for an in-depth exploration of complex phenomena within real-world contexts. It provides valuable insights into the dynamics and intricacies of student engagement with GenAIbots, including specific interactions with ChatGPT and Bing Chat.

**Participants**

As the concept of using as ChatGPT or Bing Chat as an object-to-think-with is relatively new and innovative, it has proven challenging to find teachers who are both willing and prepared to incorporate it into their classrooms for experimental purposes. The investigation did not involve any real students. Instead, the primary researcher and his graduate student, who was pursuing his Doctor degree in Science, Technology, and Engineering and acting as STEM teachers at a Brazilian secondary school acted as students in the following interaction sessions. As STEM teachers, they provided valuable insights from an instructional perspective. However, to preserve privacy, they will be randomly identified as P1 and P2.

**Procedure**

The goal with GenAIbots is not to pose a series of disconnected questions for which they provide ready-made answers. Instead, using GenAIbots as objects-to-think-with is to facilitate deeper understanding by engaging students in reflection and discovery on a given topic. This process resembles a philosophical dialogue, involving a dynamic sequence of interactions with real-time feedback, where each response prompts further user reflection and problem-solving abilities and cultivates a more profound comprehension.

Compared to ChatGPT, Bing Chat does not retain conversations beyond the duration of each chat. This means there is no easy way to preserve a record of the students' sequence of interactions with prompts and responses for subsequent analysis. Furthermore, Bing Chat does not inform its version number or release date.

The participants conducted a few sessions where they simulated STEM students' learning experiences by interacting with ChatGPT and Bing chat. Those GenAIbots facilitated personalised instruction, problem-solving support, and reflection on STEM concepts.

Based on the previous discussion on prompting, our approach was influenced by the idea that the most effective way to learn with AI is by treating it as a partner and a teacher (Mollick, 2023). To optimise our interactions, we experimented with various prompts, providing feedback to the GenAIbots and challenging them to improve. Eventually, we settled on the prompt in the $2^{nd}$ session, recognising it may not be the best choice.

**Instruments**

The study employed two primary instruments for data collection:

- GenAIbot interaction logs: Logs from participants' interactions with GenAIbots were collected, recording their sequences of interactions with prompts and responses, allowing for later analyses of the content, quality, and nature of the exchanges.
- Reflective journals: Participants maintained reflective journals throughout the study, documenting their experiences, challenges, and learning outcomes when using a GenAIbot in their STEM learning activities.

**Data Analysis**

Initially, researchers familiarised themselves with the data through repeated readings of the GenAIbots interaction logs and reflective journals. Following this, those materials were analysed to identify recurrent themes and patterns, thereby elucidating the perceived impact of GenAIbots as an object-to-think-with on STEM learning. Finally, member checking involved sharing preliminary findings among participants, allowing them to confirm, refine, or challenge each other researchers' interpretations.

**RESULTS AND ANALYSES**

This section presents GenAIbots interaction logs in summarised form (the full versions of the logs can be accessed on figshare[3]). Moreover, analyses the findings from the sessions where participants simulated STEM students' learning experiences by engaging with ChatGPT and Bing Chat.

This study aimed to investigate the effectiveness of ChatGPT and Bing Chat in enhancing STEM learning. Empathy was incorporated into the research methodology to enable the researchers to act like students and understand their perspectives.

**1st session – mass and volume**
GenAIbot: Bing Chat
Model: GPT-4 (?)
Release date: not available
Participant: P1
Usage time: 2023-Apr-22.
**Prompts and generated texts** (translated from Portuguese for convenience)
The study explored the intersection zone between the alternative conceptions of students and teachers and their impact on meaningful learning. Therefore, the aim here was to learn from the GenAIbot and not find answers on how to teach.

*P1: Good afternoon! My interest is to conduct a pre-test that will serve as a diagnosis regarding the students' alternative conceptions. The subject: Matter. Do you have any ideas?*

*Bing Chat: I understood. A diagnostic pre-test is a vital tool to assess students' alternative conceptions. The construction of a diagnostic tool can be based on alternative concepts already identified in the literature by several researchers regarding chemistry, physics, biology, and geology concepts. Here is an article that may help you better understand the subject: https://www.scielo.br/j/er/a/RmGBrVJVsnBFgVhrqv7MjXg/?format=pdf. Hope it helps!*

Bing Chat was informed of the desire to perform a diagnostic pre-test on the student's concepts. The response was positive, and it was viewed as a valuable tool for evaluating the students' concepts. The topic suggested was "Matter and its Definition." However, the article provided by Bing Chat, titled "Alternative Concepts and Biology Teaching: How to Use Differentiated Strategies in Initial Teacher Training," was not deemed interesting or motivating for the researcher and much less for a student.

---

[3] https://doi.org/10.6084/m9.figshare.22723862.v1

*P1: I would like to start a dialogue here regarding the students' conceptions on the subject: Matter. I am interested in knowing how this dialogue between you and me would proceed. Could it be?*

*Bing Chat: Sure! We can talk about it. What would you like to know about the students' conceptions regarding Matter?*

*P1: For example: explain in your own words what the words mean: mass and volume. Let´s start?*

*Bing Chat: Sure! Mass is the amount of matter an object has. It is a property that does not depend on the object's location and is measured in kilograms (kg). Volume is the space occupied by an object. It is a property that can change depending on the object's location and is measured in cubic meters ($m^3$). I hope this helps! :)*

*P1: I believe that the concepts you provided do not match the students' conceptions. Acting as a student, what would your answer be?"*

*Bing Chat: I understand. As a student, I would say that mass is the weight of an object, and volume is the object's size. These are students' most common alternative conceptions of these concepts. I hope this helps.*

It was unclear whether Bing Chat was able to distinguish the proficiency level of the interlocutor. To address this, a request was made for a dialogue between the researcher and Bing Chat. The researcher then asked for the definition of mass and volume, but the answer given was the same as found in most textbooks and websites, which did not match students-level conceptions. The researcher informed this to Bing Chat and asked for its conception as a student. Bing Chat then exchanged the concept of mass for weight and volume for the size of the object, reflecting the usual student's understanding.

*P1: Oh, now it's more like the student's conceptions. Continuing with this idea, you being a student, could you indicate two situations that happen in real life for mass and volume?*

*Bing Chat: Sure! One situation where mass is essential is when you need to know the weight of an object. For example, if you're shopping for fruit at the supermarket, you might want to know the weight of the fruit, so you know how much you'll pay. One situation where volume is essential is when you need to know how much space an object takes up. For example, if you buy a box to store your stuff in, you might want to know the box's volume to know if your stuff will fit. Hope this helps!*

The researcher then requested day to day examples that related to mass and volume, but Bing Chat again related mass to weight and volume to the space occupied by an object.

*P1: I didn't understand the examples for mass and volume you provided. To know the mass and volume of a fruit, would it be enough then to weigh it? Could you provide counterexamples?*

*Bing Chat: I'm sorry for the confusion. In fact, the weight of an object is a measure of the gravitational force acting on it and is measured in Newton (N). Mass is a property that does not depend on the object's location and is measured in kilograms (kg). An example of a situation where mass is important is when you need to know how much matter an object has. For example, if you're making a cake recipe, you'll need to know how much flour you're going to use. For this, you can measure the flour mass. An example of a situation where volume is important is when you need to know how much space an object occupies. For example, if you're buying a box to hold your stuff, you might want to know the box's volume so you know your stuff will fit. I hope I clarified your doubts!*

*P1: Acting like a student, now I don't understand. gravitational force, newtons, what mass is a property that does not depend on where the object is. But how to measure the mass of flour?*

*Bing Chat: To measure the flour mass, you can use a digital scale and weigh the amount of flour you need for your recipe. The unit of measurement for mass is the kilogram (kg) or the gram (g). Another option is to use a measuring cup or spoons to measure the amount of flour. However, it's important to remember that the density of flour can vary depending on the type and brand, which can affect how much flour you need for your recipe.*

*P1: I think it's getting complicated to understand about mass and volume. Does it mean that I can measure the flour mass using a cup? You mentioned the density of the flour. What did you mean?*

*Bing Chat: Yes, you can measure flour mass using measuring cups or spoons. However, it's important to remember that the density of flour can vary depending on the type and brand, which can affect how much flour you need for your recipe. Density is a physical property that relates the mass of an object to its volume. It is measured in kilograms per cubic meter (kg/m³) or grams per cubic centimeter (g/cm³).*

The interaction between Bing Chat as a student and the absence of a teacher during the class may bring problems for learning. It is like the spread of fake news, where students copy and paste incorrect or incomplete concepts, making it difficult for teachers to correct them later. The researcher requested counterexamples when Bing Chat gave examples that reinforced scientifically accepted concepts of mass and volume. Bing Chat then provided examples that reflected the student's alternative conceptions, but also includes the new concept of "density."

***P1:*** *The subject I began to discuss with you was about matter. But due to your suggestion, which was an article, I opted for the definition of mass and volume. Then you mention gravitational force, newtons, and density. After all, what's your deal with matter, mass, and volume?*

***Bing Chat:*** *Matter is anything that has mass and takes up space. Mass is a property of matter that measures the amount of matter an object has and is measured in kilograms (kg) or grams (g). Volume is a property of matter that measures the space an object occupies and is measured in cubic meters (m³) or liters (L).*

***P1:*** *What measurements would I use for olive oil? For the rice? For the tomato? For flour? For the garlic? For milk? For coffee?*

***Bing Chat:*** *To measure olive oil, you can use a tablespoon or measuring cup. For eggs, you can use a digital scale or measuring cup. For rice, you can use a measuring cup or a digital scale. For tomatoes, you can use a digital scale or measuring cup. For flour, you can use a digital scale or measuring cups and spoons. For garlic, you can use a digital scale or teaspoons. For milk, you can use measuring cups and spoons or a digital scale. For coffee, you can use tablespoons or measuring cups.*

***P1:*** *But after all, what would be the measurements of mass, and which would be the measurements of volumes.*

***Bing Chat:*** *Mass measurements include kilograms (kg), grams (g), and milligrams (mg). Volume measurements include cubic meters (m³), liters (L), milliliters (ml) and cubic centimeters (cm³).*

The researcher then tried to return to the initial questions, but it was noted that the formulation of questions left room for various interpretations regarding measurement. The researcher reformulated the question to reflect the most common units of measurement for mass and volume. However, Bing Chat cited liter and milliliter as their units, in addition to m³ and cm³, which are less common in everyday life.

**2rd session – Elevator Affects Apparent Weight.**
GenAIbot: ChatGPT
Model: GPT-4
Release date: March 23
Participant: P2
Usage time: 2023-Apr-28.

***P2:*** *I am about to embark on a dialogue where I will embody the role of a student, engaging with you, ChatGPT, as an object-to-think-with, with the primary goal of delving into and deliberating upon physical [or any chosen theme] concepts. As an object-to-think-with, I kindly request your support in furnishing engaging and personalized responses to the student's inquiries. It is crucial that our interaction adopts a Socratic approach, encouraging the student to reflect upon their existing knowledge, stimulating critical thinking, and aiding in their refinement of comprehension rather than offering ready-made answers. Are you prepared to embark on this collaborative and thought-provoking journey?*

***ChatGPT:*** *Yes, I am prepared to embark on this collaborative and thought-provoking journey with you. As an object-to-think-with, I will strive to engage you in a Socratic dialogue, fostering reflection, critical thinking, and a deeper understanding of the chosen theme. Please feel free to present your initial inquiry or topic, and we can begin our exploration together.*

***P2:*** *I don't get weight. When I take the elevator down, I feel less weight, and when I take it up, I feel more.*

***ChatGPT:*** *That's an interesting observation you've made! Let's try to understand this phenomenon together. Can you tell me what you know about weight and how it's related to the force of gravity?*

In the context of taking an elevator, the student's question is grounded in a real-life scenario, which makes it relatable and engaging. However, the student seems to have a misconception about the concept of weight and the sensation of weightlessness. ChatGPT has apparently taken over its role as an object-to-think-with by posing questions to the student and encouraging critical thinking and active engagement in the learning process. Furthermore, ChatGPT builds upon the student's prior knowledge by inquiring about their existing understanding of weight and gravity, facilitating a meaningful connection between new and existing information.

***P2:*** *All I know is that weight is the force of gravity acting on an object.*

***ChatGPT:*** *Great, that's a good starting point! Weight is indeed the force of gravity acting on an object, and it depends on both the mass of the object and the acceleration due to gravity. Now, let's consider what happens when you're in an elevator. Can you describe the forces acting on your body when the elevator is at rest, accelerating upwards, and accelerating downwards?*

Again, ChatGPT pose a question to the student rather than offering ready-made answers, therefore encouraging critical thinking and active engagement in the learning process.

*P2: I know there's the weight force pulling me down, but as I said, I feel less weight when I'm going down and more weight when I'm going up than when I'm standing on the ground.*

*ChatGPT: That's correct. Your weight, which is the force of gravity pulling you downwards, is always acting on you. However, when you're in an elevator, there's another force at play, which is the force exerted by the elevator floor on your body. This force is called the normal force. When the elevator is at rest or moving at a constant speed, the normal force is equal to your weight, and you feel your "usual" weight. However, when the elevator accelerates, the situation changes.*

1. *When the elevator accelerates upwards, the normal force needs to counteract both your weight and the additional force required for upward acceleration. As a result, the normal force becomes greater than your weight, and you feel "heavier."*

2. *When the elevator accelerates downwards, the normal force only needs to counteract the difference between your weight and the downward acceleration. This means the normal force is now less than your weight, making you feel "lighter."*

*Does this explanation help you understand why you feel different weights when the elevator is moving up or down?*

ChatGPT's answer provides a clear and concise explanation, differentiating between the actual weight, which is the force of gravity acting on an object, and the apparent weight due to the elevator's acceleration. The response may help the student understand the physics behind the sensation of weightlessness in an elevator and its connection to acceleration, expanding their knowledge and grasp the intricate nature of these concepts.

*P2: But I don't really feel that force of gravity acting on my body, just that force that the elevator floor exerts upwards.*

*ChatGPT: You're right in that what you're feeling is not the force of gravity directly, but rather the normal force exerted by the elevator floor on your body. The force of gravity is an invisible force that acts between objects with mass, but we don't sense it directly. Instead, we sense the normal force, which is the force exerted by a surface in response to the force of gravity. In everyday situations, like standing on the ground or on the floor of an elevator, the normal force is what we perceive as our weight. It's the force that prevents us from sinking into the ground or the elevator floor due to gravity. When the elevator accelerates, as we discussed earlier, the normal force changes, and we perceive these changes as fluctuations in our apparent weight.*

The student seems to maintain his misconception about the concept of weight and weightlessness. ChatGPT reiterates that, in an accelerating elevator, the force of gravity is still pulling the person towards the Earth, but the elevator's acceleration adds an extra force that affects their apparent weight. It maintains the distinction between actual weight and apparent weight and emphasizes once again that the actual weight does not change since the force of gravity remains constant. Yet, as an object-to-think-with, ChatGPT could have built on this knowledge by using familiar examples and discussing other situations where similar sensations are experienced (e.g., riding a roller coaster or a car accelerating/decelerating).

**P2:** *I understand that the origin of the force of gravity is the Earth but what is the origin of that 'normal' force?*

**ChatGPT:** *That's a great question! The origin of the normal force lies in the interactions between the atoms and molecules in the objects involved. In the case of standing on the elevator floor or any surface, the normal force originates from the electromagnetic forces between the atoms and molecules in both your body and the surface you're standing on. When you stand on a surface, the atoms and molecules in your body exert a force on the atoms and molecules in the surface due to gravity. In response, the atoms and molecules in the surface exert an equal and opposite force on your body. This opposing force is the normal force, and it acts perpendicular to the surface. It is the result of the repulsive electromagnetic interactions between the negatively charged electrons in the atoms of your body and those of the surface. The normal force is what keeps you from sinking into the surface or falling through it. When you feel your weight or experience changes in it due to an elevator's acceleration, you are actually sensing the normal force acting on your body.*

In this dialogue, the student inquires about the origin of the normal force, a crucial concept for comprehending the forces acting on objects in contact with surfaces. ChatGPT offers a concise explanation, highlighting that the normal force arises from electromagnetic interactions between atoms and molecules present in both the object and the surface it touches. By relating the abstract notion of the normal force to tangible everyday scenarios, such as perceiving one's weight or sensing changes in it due to an elevator's acceleration, ChatGPT assists the student in establishing connections and facilitates a deeper understanding of the concept. This dialogue effectively showcases the use of ChatGPT as an object-to-think-with, guiding the student through intricate physical concepts and encouraging further exploration by relating them to real-life experiences.

**P2:** *Your earlier answer about added force and acceleration on the elevator confused me. Too many things acting at once.*

***ChatGPT:*** *I apologize if my previous explanation was confusing. Let's break it down into simpler terms and focus on the forces acting on your body when you're in an elevator.*

1. *At rest or moving at a constant speed: When the elevator is not accelerating, there are two forces acting on your body: the force of gravity pulling you downward (your weight) and the normal force exerted by the elevator floor pushing you upward. In this case, the normal force is equal in magnitude to your weight but opposite in direction, so you feel your "usual" weight.*

2. *Accelerating upwards: When the elevator accelerates upward, the normal force needs to provide not only the force to counteract your weight but also an additional force to make you accelerate along with the elevator. As a result, you feel a greater normal force acting on your body, making you feel "heavier."*

3. *Accelerating downwards: When the elevator accelerates downward, the normal force now only needs to counteract the difference between your weight and the downward acceleration. This means that the normal force acting on your body is less than your weight, making you feel "lighter."*

*The key takeaway is that what you feel as your "weight" in the elevator is actually the normal force exerted by the elevator floor on your body. This normal force can change depending on the acceleration of the elevator, causing you to feel heavier or lighter.*

*I hope this explanation is clearer.*

ChatGPT acknowledges the student's confusion and simplifies the breakdown of forces in different scenarios: at rest, accelerating upwards, and accelerating downwards. This adaptive approach helps the student understand the forces involved and their impact on weight perception. The interaction showcases ChatGPT's ability to adjust explanations, fostering clearer comprehension and facilitating an easier grasp of the concept.

***P2:*** *Hmmm. My physics teacher also talked about this normal force. He said that a table exerts an upward force on a book on top of it. But the table is inanimate, how can it exert force?*

***ChatGPT:*** *It's true that a table is an inanimate object, but it can still exert a force due to the properties of its materials and the physical laws that govern the interactions between objects. In this case, the normal force exerted by the table on the book is a result of the electromagnetic interactions between the atoms and molecules in the table's surface and those in the book. When you place a book on a table, the force of gravity pulls the book downward, causing the atoms and molecules in the book to come into close contact with those in the table's surface. The negatively charged electrons in the atoms of the book and the table repel each other through electromagnetic*

*forces. These repulsive forces prevent the book from sinking into the table and act in the opposite direction of gravity, effectively creating the normal force that holds the book up. Even though the table is inanimate, the fundamental forces between the atoms and molecules within it are still at play. This is what allows the table to exert an upward force, the normal force, on the book.*

*I hope this explanation helps clarify how an inanimate object like a table can exert a force on another object.*

In this exchange, ChatGPT explains that the normal force is a result of electromagnetic interactions between the atoms and molecules in the table and the book, emphasizing that it is not a conscious action of the table. By focusing on electromagnetic interactions, ChatGPT helps the student grasp the concept of the normal force and forces between inanimate objects and prompts deeper reflection on forces in everyday situations.

***P2:*** *He also said that for a body to be at rest, the resultant of the forces must be zero. So, how does the table know exactly what value it must make of this normal force to compensate for the weight of the book?*

***ChatGPT:*** *That's an excellent question! The table doesn't "know" the weight of the book or what value of the normal force it must exert to balance it. Instead, this balancing of forces occurs naturally due to the physical properties and laws governing the interaction between the book and the table. When you place a book on a table, the force of gravity pulls the book downward. This causes the atoms and molecules in the book to come into close contact with those in the table's surface. The repulsive electromagnetic forces between the atoms in both objects increase as they get closer together. The table doesn't need to "know" or calculate the force; the normal force automatically adjusts based on the interactions between the atoms and molecules. Once the repulsive forces between the book and table become equal in magnitude to the force of gravity acting on the book, the book will come to rest. At this point, the normal force exerted by the table on the book is equal to the weight of the book, and the net force on the book is zero. Since there is no net force acting on the book, it remains at rest, in accordance with Newton's first law of motion. So, it's not a matter of the table knowing what force to exert, but rather a natural response arising from the physical properties and laws governing the interactions between the book and the table.*

ChatGPT explains how this balance of the normal force to the weight of the book occurs naturally through the physical properties and laws governing the interaction between the book and the table, rather than the table possessing knowledge or performing calculations. The explanation enables the student to grasp the concept of equilibrium and the involved forces, rectifying the misconception and fostering a more accurate mental model of these phenomena.

*P2: Does it mean that the tabletop works like a spring that deforms according to the force exerted on it?*

*ChatGPT: Yes, that's a good way to think about it! In fact, the tabletop, as well as any solid object, behaves like a very stiff spring when a force is applied to it. This behavior is due to the elastic properties of the material, which arise from the atomic and molecular interactions within the material. When you place a book on a tabletop, the force of gravity causes a slight deformation or compression in the table's surface. This compression is typically too small to see with the naked eye, but it is present, nonetheless. The atoms and molecules in the table's material are pushed closer together, which increases the repulsive electromagnetic forces between them. These forces oppose the compression, acting like the restoring force in a spring. As a result, the table exerts an upward force on the book, which we call the normal force. The normal force adjusts in magnitude according to the weight of the book, just like a spring's restoring force changes with the amount of compression or elongation. So, in a sense, the tabletop does work like a spring that deforms according to the force exerted on it, even though the deformation is generally too small to observe.*

ChatGPT's explanation enhances the student's comprehension that the normal force adjusts in magnitude relative to the weight of the book, like how a spring's restoring force changes with compression or elongation. By validating the student's analogy and providing a clear explanation, ChatGPT as an object-to-think-with reinforces the student's comprehension of the concept and underscores the use of familiar analogies, such as springs, to understand physical interactions between objects.

*P2: But back to the subject of weight, what about space station astronauts without weight? There are no normal forces there, right?*

*ChatGPT: You're correct that astronauts on a space station experience weightlessness, which means they don't feel the normal force like they would on Earth. However, it's important to note that the force of gravity is still acting on the astronauts, even though they don't feel their weight. Astronauts aboard a space station are in a state of continuous freefall, as the space station orbits Earth. They're essentially falling towards Earth at the same rate that the space station is falling. This is due to the fact that the space station's horizontal velocity is balanced with the gravitational force pulling it towards Earth, which causes the space station to follow a curved path around the planet. Since the astronauts are falling along with the space station, there is no surface pushing up against them to create a normal force. As a result, they don't feel their weight, and they*

*experience weightlessness, also known as microgravity. So, you're right that there are no normal forces acting on astronauts in a space station, but it's important to understand that this weightlessness is due to their state of continuous freefall rather than the absence of gravity.*

In this interaction, with ChatGPT the student to delves into weight, normal force, and microgravity within the context of space station astronauts. ChatGPT's explanation aids the student's comprehension that the weightlessness experienced by astronauts in orbit stems from the absence of normal forces rather than gravity itself.

*P2: What do you mean by freefall? Are they falling like an elevator with a broken cable?*

*ChatGPT: Yes, in a way, astronauts aboard a space station are in a state similar to that of freefall, like being in an elevator with a broken cable. However, the key difference is that astronauts in a space station are also moving horizontally at a high speed, which keeps them in a stable orbit around Earth. When an object is in freefall, it means that the only force acting on it is gravity, causing it to accelerate towards the center of the Earth. In the case of a space station, it is indeed falling towards Earth due to gravity, but at the same time, it is also moving horizontally at a very high speed (around 28,000 km/h or 17,500 mph). This high horizontal speed causes the space station to move forward at the same rate that it is falling towards Earth. The result is that the space station follows a curved path around the Earth, essentially falling around the planet. Since the space station is in continuous freefall, the astronauts inside it are also in freefall. With no surface to exert a normal force on them, they don't feel their weight and experience weightlessness, or microgravity. So, while astronauts are indeed in a state of freefall, it is their horizontal velocity that keeps them from falling straight down to Earth, instead maintaining a stable orbit.*

*I hope this explanation helps you understand the concept of freefall and orbiting in the context of a space station.*

This interaction enhances the student's comprehension of the term "freefall" and its connection to weightlessness for space station astronauts. The explanation underscores the significance of horizontal velocity in maintaining a stable orbit and highlights the disparity between freefall in the context of a space station and other scenarios like a falling elevator.

*P2: You said that in space, the gravitational force acting on an object is greatly reduced, but it is not the gravitational force that keeps the Moon circling around the Earth?*

*ChatGPT: I apologize for any confusion in my previous explanations. In fact, the force of gravity is what keeps the Moon orbiting around the Earth. The gravitational force between the Earth and the Moon is not greatly reduced; rather, the gravitational force is weaker at the Moon's distance*

*from Earth compared to the surface of the Earth. The Moon orbits the Earth due to the balance between its inward gravitational force, which pulls it towards the Earth, and its outward centrifugal force, which results from its orbital motion around the Earth. The Moon's gravitational attraction to Earth provides the centripetal force that keeps it moving in a curved path, while its orbital velocity prevents it from falling straight into Earth. Similarly, astronauts aboard a space station are also subjected to the force of gravity. However, they experience weightlessness because they are in a state of continuous freefall, as the space station orbits Earth. The gravitational force in low Earth orbit, where space stations typically reside, is only slightly weaker than on Earth's surface (about 90% of the surface gravity). In summary, the force of gravity does play a crucial role in keeping the Moon orbiting around the Earth and is still present in space, albeit slightly weaker than on Earth's surface. Weightlessness experienced by astronauts in a space station is due to their continuous freefall state while orbiting Earth.*

In this part of the dialogue, ChatGPT corrects the previous statement, focusing on the balance between the Moon's gravitational force and its centrifugal force resulting from its orbital motion, clarifying the factors that contribute to the Moon's orbit around Earth. This clarification helps dispel confusion and reinforces the understanding of weightlessness experienced by astronauts in a space station as a result of continuous freefall while still being influenced by gravity.

**P2:** *You spoke of centrifugal force. But isn't centrifugal force a fictitious force?*

**ChatGPT:** *You're correct that the centrifugal force is often referred to as a "fictitious force" or "pseudo-force." It is not a real force in the sense that it doesn't arise from a direct interaction between objects, like gravitational or electromagnetic forces. Rather, the centrifugal force is a result of the inertia of an object moving in a curved path, and it is only experienced in a rotating or non-inertial reference frame. When we observe a rotating system from a non-rotating (inertial) frame of reference, we don't need to consider centrifugal force. Instead, we can explain the motion of objects using centripetal force, which is the real force that acts towards the center of the circular path, causing the object to move in a curve. In the case of the Moon orbiting the Earth, the gravitational force provides the necessary centripetal force to maintain its curved path. However, when we analyze the motion from the perspective of a rotating (non-inertial) frame of reference, like an observer on the Moon, we need to introduce the concept of centrifugal force. This force appears to act away from the center of the circular path and is equal in magnitude but opposite in direction to the centripetal force. So, while the centrifugal force is a fictitious force, it can still be a useful concept when examining problems from a rotating frame of reference. In the context of the Moon orbiting the Earth, it's the balance between the gravitational force*

(providing centripetal force) and the Moon's orbital motion that maintains the Moon's stable orbit.

ChatGPT acknowledges the student's questioning and clarifies that the centrifugal force is indeed a fictitious force, arising from the observation of motion in a rotating reference frame. By distinguishing between real and fictitious forces and emphasizing the significance of selecting the appropriate frame of reference, the explanation helps deepen the student's understanding of the concept of centrifugal force and its relevance in analysing rotational motion.

*P2: My physics teacher said that if I thought centrifugal force was fictitious i should open the door on my side of the car when it was making a curve.*

**ChatGPT:** *Your physics teacher's suggestion is likely intended to help you appreciate the effects of the centrifugal force, even though it is a fictitious force. When you're inside a car that's turning, you're in a non-inertial (rotating) frame of reference, and you may feel like you're being pushed towards the outer side of the curve. This apparent outward push is the effect of the centrifugal force. If you were to open the car door while the car is turning, the effect of the centrifugal force might cause you to slide or be pushed out of the car. From an inertial frame of reference (e.g., someone observing the car from a distance), there is no centrifugal force acting on you. Instead, the observer would see that you're moving along a curved path due to the centripetal force (friction between the tires and the road), which keeps the car moving in the curve. Your body, due to its inertia, wants to continue moving in a straight line, but the car's interior (the door, seat, etc.) provides the necessary force to keep you moving along the curved path with the car. So, while the centrifugal force is indeed fictitious and only appears in a rotating or non-inertial frame of reference, its effects can still be felt by someone inside the car. The example your teacher provided is aimed at helping you understand the practical implications of the centrifugal force, even though it is not a "real" force in the same way as forces like gravity or friction.*

By addressing the teacher's comment and emphasizing the importance of understanding practical implications, ChatGPT provides insights into the concept of the centrifugal force and the role of the frame of reference in perceiving forces.

*P2: OK. I may have grasped the weight idea a little bit. Thanks.*

**ChatGPT:** *You're welcome! I'm glad I could help clarify the concept of weight for you. Understanding weight, as well as related forces and phenomena like gravity, normal force, and centrifugal force, is essential for building a strong foundation in physics.*

*If you have any more questions on this topic or any other subjects, please don't hesitate to ask. I'm here to help you with any doubts or inquiries you may have. Remember, asking questions and*

*engaging in discussions is an excellent way to learn and deepen your understanding. Good luck with your studies!*

This session accurately captured the role of ChatGPT in guiding the student through a learning journey to address their initial difficulty in understanding the concept of weight. By acting as an object-of-think-with, ChatGPT was able to facilitate the student's learning process in a meaningful and engaging manner. The exploration covered various aspects of the student's question, helping them gain a deeper understanding of related concepts such as electromagnetic forces, elastic properties of materials, normal forces, acceleration, inertia, and the sensation of weightlessness.

It is crucial to highlight that, upon comparing the final session to the two preceding ones, the initial prompting appears to have 'programmed' ChatGPT more effectively, enabling it to serve as an object-to-think-with. This observation underscores the significance of carefully crafted prompts in optimizing the AI model's performance within the context of the experiment.

Despite the similarities between ChatGPT and Bing Chat, their performance in this experiment varied significantly. It appears that when operating in Portuguese, Bing Chat relied on lower-quality internet sources, which adversely affected the quality of its responses. In contrast, ChatGPT maintained the use of sources from its training data, which were seemingly more meticulously curated, resulting in better responses.

**DISCUSSIONS**

Overall, the success of this learning journey with GenAIbots acting as objects-to-think-with showcase several effective pedagogical strategies that can facilitate a deeper understanding of the concepts involved. Firstly, positive reinforcement is employed as GenAIbots acknowledge the student's observation as interesting, validating their experience and nurturing their curiosity. Additionally, a Socratic-like questioning approach is utilized, as opposed to directly providing an answer. By posing questions to the student, GenAIbot encourages critical thinking and active engagement in the learning process. Furthermore, GenAIbots build upon the student's prior knowledge by inquiring about their existing understanding of weight and gravity, facilitating a meaningful connection between new and existing information. Lastly, the response maintains a student-centred learning approach, focusing on the student's thoughts and experiences, and fostering an environment conducive to exploration, discussion, and reflection. By encouraging the student to actively engage in the learning process, GenAIbot can create a more enriching and personalized educational experience, as expected from an object-to-think-with.

The interactive nature of this session allowed the student to progressively build on their existing knowledge, while GenAIbot provided relevant explanations, examples, and analogies to support their learning. As an object-of-think-with, GenAIbots were able to adapt its responses to the student's questions and confusion, providing tailored explanations and addressing misconceptions. This demonstrates the potential of AI-driven tools like GenAIbot in promoting active learning, personalized instruction, and conceptual understanding in various domains, including physics.

This study highlights the value of using AI to enhance educational experiences and concluded that AI-driven chatbots, such as ChatGPT, Bing Chat, and other generative models, can serve as innovative laboratories for exploring a wide range of scenarios that often pose challenges to students' learning. As previously mentioned, employing well-crafted sentences, a technique known as prompt-crafting (Mishra et al., 2023), can help minimize the risk of misinterpretation and support the assessment of students' alternative conceptions in the context of scientific knowledge.

The process of crafting a prompt that elicits the desired response is a crucial aspect of the iterative reflections and interactions that occur when using generative AI-powered chatbots (GenAIbots) as objects-to-think-with. This approach encourages students to think critically and refine their understanding of various concepts while actively engaging with the AI-powered tool.

Additionally, the study emphasized the importance of providing comprehensive training to educators before integrating these tools into their classrooms. Proper training enables teachers to effectively guide students in articulating their thoughts and constructing well-reasoned arguments. This preparation allows educators to evaluate the performance of GenAIbots when confronted with relevant and convincing counterexamples and fosters fruitful discussions among colleagues.

By harnessing the capabilities of generative AI chatbots as learning tools, educators can create more engaging and personalized learning experiences, ultimately fostering deeper comprehension of complex concepts and promoting critical thinking in students across diverse subject areas.

**CONCLUSION**

In conclusion, our results show that GenAIbots, including ChatGPT and Bing chat, are powerful objects-to-think-with in STEM education, offering numerous advantages for cognitive development and learning. These benefits encompass interactivity, personalization, multidisciplinary knowledge, critical thinking, problem-solving, creativity encouragement, and collaboration. Users can engage in dynamic conversations, explore tailored content, access diverse knowledge, develop critical skills, and stimulate creativity. Furthermore, GenAIbots facilitate group discussions and collaboration, deepening topic understanding.

GenAIbots provide interactive and personalized learning experiences, allowing users to engage in dynamic conversations, ask questions, and receive instant feedback. They can adapt to individual needs and interests, making learning more engaging and relevant for users. With extensive training data, GenAIbots offer a wide range of knowledge across various disciplines, supporting users in expanding their understanding of different subjects.

By interacting with GenAIbots, users can develop critical thinking, problem-solving, and creativity skills. They can articulate thoughts, ask probing questions, and analyse information provided by the AI. In group settings, GenAIbots facilitate discussions, debates, and collaborative projects, helping users share knowledge and gain deeper insight into topics.

In line with Seymour Papert's concept, GenAIbots serve as exceptional objects-to-think-with, fostering a more engaging, interactive, and inclusive learning environment. They promote creativity, collaboration, exploration, critical thinking, problem-solving, and a deeper understanding of subject matter.

Incorporating GenAIbots into STEM education aligns with constructionist learning theory, which emphasizes hands-on, experiential learning and cognitive tools. By utilizing GenAIbots, students can hone their problem-solving, critical thinking, and creative skills in a more accessible and engaging learning environment, empowering them to succeed in STEM fields.

**Limitations of the study and possible future studies**

While this study faced limitations due to its single case study design, valuable insights emerged from the extensive interactions between students and GenAIchats. Despite constraints like lack of generalizability and potential biases, the in-depth exploration uncovered the system's untapped potential. Crafting good prompts is an area for future work, experimenting with different types and examining how GenAIchat's performance as an object-to-think-with improves with specific prompts like the following the one used in the 2nd session above. Implementing GenAIbots in STEM learning requires weighing benefits and limitations, ensuring accurate information and considering reduced human interaction. Integrating GenAIbots with other educational tools or activities that promote collaborative dialogue among learners may address these concerns.